\documentclass[journal]{IEEEtran}

\usepackage{graphicx}
\usepackage{epsfig}
\usepackage{epstopdf}

\usepackage[cmex10]{amsmath}
\usepackage{amssymb}
\usepackage{bm}
\usepackage{cite}
\usepackage{amsmath}
\usepackage{wasysym}
\usepackage{booktabs}
\usepackage{paralist}
\usepackage{balance}
\usepackage{subfigure}
\usepackage{multicol}
\usepackage[ruled,linesnumbered]{algorithm2e}
\usepackage{amsmath, amssymb,amsthm}
\usepackage{upgreek}

\newtheorem{proposition}{Proposition}
\newtheorem{lemma}{Lemma}

\ifCLASSINFOpdf
\else
\fi

\begin{document}
%
\title{Load Coupling Power Optimization in Cloud Radio Access Networks}
%
%
%

\author{Qiang~Fan,
        Hancheng~Lu,~\IEEEmembership{Member,~IEEE,}
        Wei~Jiang,
        Peilin~Hong,~\IEEEmembership{Senior Member,~IEEE,}
        Jun~Wu,~\IEEEmembership{Senior Member,~IEEE,}
        and~Chang Wen Chen,~\IEEEmembership{Fellow,~IEEE}
\IEEEcompsocitemizethanks{\IEEEcompsocthanksitem Qiang Fan, Hancheng Lu, Wei Jiang and Peilin Hong are with the Department
of Electrical Engineering and Information Science, University of Science and Technology of China, Hefei,
China (Email: fanq@mail.ustc.edu.cn, hclu@ustc.edu.cn, jweiseu@mail.ustc.edu.cn, plhong@ustc.edu.cn).
\IEEEcompsocthanksitem Jun Wu is with the Computer Science and Technology Department, Tongji University, Shanghai, China (Email: wujun@tongji.edu.cn).
\IEEEcompsocthanksitem Chang Wen Chen is with the State University of New York, Buffalo, NY 14260 USA (Email: Chencw@buffalo.edu).}
\thanks{This work was supported in part by the National Science Foundation of China (No.61390513, No.91538203) and the National High Technology Research and Development Program of China (863 Program) (No.2014AA01A706).}
}

\maketitle

\begin{abstract}
Recently, Cloud-based Radio Access Network (C-RAN) has been proposed as a potential solution to reduce energy cost in cellular networks. C-RAN centralizes the baseband processing capabilities of Base Stations (BSs) in a cloud computing platform in the form of BaseBand Unit (BBU) pool. In C-RAN, power consumed by the traditional BS system is distributed as wireless transmission power of the Remote Radio Heads (RRHs) and baseband processing power of the BBU pool. Different from previous work where wireless transmission power and baseband processing power are optimized individually and independently, this paper focuses on joint optimization of allocation for these two kinds of power and attempts to minimize the total power consumption subject to Quality of Service (QoS) requirements from users in terms of data rates. First, we exploit the load coupling model to express the coupling relations among power, load and user data rates. Based on the load coupling mode, we formulate the joint power optimization problem in C-RAN over both wireless transmission power and baseband processing power. Second, we prove that operating at full load may not be optimal in minimizing the total power consumption in C-RAN. Finally, we propose an efficient iterative algorithm to solve the target problem. Simulations have been performed to validate our theoretical and algorithmic work. The results show that the proposed algorithm outperforms existing schemes (without joint power optimization) in terms of power consumption.
\end{abstract}

\begin{IEEEkeywords}
Cloud-based radio access network, load coupling model, power optimization.
\end{IEEEkeywords}

%
\IEEEpeerreviewmaketitle

\section{INTRODUCTION}
%
%
%
%
With the development of wireless technologies and popularization of smart phones, mobile traffic grown 4000-fold over the past 10 years and is expected to continue its growth at a compound annual growth rate of $53$ percent from $2015$ to $2020$ \cite{Cisco}. The resulting problem of energy consumption on the Information and Communications Technology (ICT) has become a serious issue. The statistics show that ICT consumes a percentage between $2\%$ and $10\%$ of the world total power. Furthermore, Radio Access Networks (RANs), which account for around $70\%$ of energy consumption in cellular networks, are considered as one of the top energy consumers of the ICT sector \cite{CMRI}. Based on the aforementioned statistics, the problem of energy saving in RANs has received increasing attention from academia and industry. Many attempts have been made to improve the energy efficiency and to reduce energy cost in RANs, among which Cloud-based RAN (C-RAN) \cite{N.Saxena}\cite{J.Wu} has been considered as one of the most promising solutions from the view point of the RAN architecture.

The most distinctive feature of C-RAN is that the baseband processing capabilities are separated from the Base Stations (BSs) and centralized in a cloud computing platform in the form of BaseBand Unit (BBU) pool. C-RAN distributes the traditional BS system into three components, i.e., Remote Radio Heads (RRHs) performing the radio function at the remote locations, BBU pool placed hundred or thousand of meters away, and backhaul links connecting the RRHs and the BBU pool \cite{Y.Cai}. By doing so, energy cost for supporting facilities such as air conditioning is reduced significantly at the remote locations. This benefit is important when more cells are deployed to meet traffic growth. Moreover, energy efficiency can be improved by sharing the baseband processing resources among RRHs \cite{M.Peng}. In the BBU pool, the number of active BBUs is varied according to the fluctuating traffic from RRHs. Therefore, C-RAN has more advantages in energy saving compared to the traditional BS system \cite{A.Checko}.

Under scenarios that the users receive various kinds of traffic, power optimization is an effective way to minimize energy consumption in RANs \cite{H.Holtkamp}\cite{C.Ho}. Different from the traditional BS system, C-RAN brings more challenges in power optimization due to the distribution of power consumption between RRHs and the BBU pool. In the traditional BS system, it has been proved that optimal power can be achieved by operating at full load at each BS \cite{C.Ho}. Here load of BS is defined as the average usage level of time-frequency resources at BS and time-frequency resources are allocated in the form of Resource Blocks (RBs) \cite{F.Z.Kaddour} in Long Term Evolution (LTE) netwoks. However, this cannot be applied to C-RAN. In C-RAN, three types of power consumption are considered, i.e., wireless transmission power of RRHs, baseband processing power of the BBU pool and backhaul transmission power of links between RRHs and the BBU pool \cite{D.Sabella}. These different types of power interact with each other, making the power optimization problem more complicated. To tackle this power optimization problem, we assume that the Quality of Service (QoS) requirement from each user is measured in terms of data rate. In this case, backhaul transmission power, which is mainly determined by the data rates of users served by RRHs, can be ignored in power optimization under given user data rates. We will focus on the other two kinds of power. Similar to that in the traditional BS system, wireless transmission power couples with load at different RRHs due to mutual interference among cells \cite{C.Ho}\cite{I.Siomina}. At the same time, baseband processing power is allocated according to the number of used RBs in proportion to load \cite{D.Sabella}. As wireless transmission power and baseband processing power interact with each other, they should be leveraged to minimize the total power consumption in C-RAN.

In previous work on addressing the energy saving problem in C-RAN, three different components, i.e., RRHs, the BBU pool and the backhaul links, perform power optimization individually and independently \cite{V.N.Ha,G.Zhai,M.Qian}. In such case, the total power of C-RAN, consisting of wireless transmission power, backhaul transmission power and baseband processing power, cannot reach optimum. Different from previous work, we focus on joint power optimization by considering all three C-RAN components in order to minimize the total power consumption of C-RAN. Without loss of generality, consider scenarios where the data rates are represented by the QoS requirements of users served by RRHs. In such scenarios, we can ignore the backhaul power and concentrate on wireless transmission power and baseband processing power in power optimization. The motivation for joint power optimization in C-RAN is based on the tradeoff between wireless transmission power and baseband processing power. At RRHs, wireless transmission power can be reduced by increasing load. On the contrary, baseband processing power at the BBU pool will be increased as more resource blocks are used. Thus, there exists optimal operating power at RRHs and the BBU pool which minimizes the total power consumption in C-RAN. Besides the aforementioned coupling relation between RRHs and the BBU pool, RRHs also couple with each other in terms of power and load due to mutual interference \cite{I.Siomina}. All of these coupling relations make the joint power optimization problem in C-RAN more challenging.

To design an efficient solution to the joint power optimization problem in C-RAN, the first step is to express the coupling relations in proper forms. Towards this end, we explore the load coupling model \cite{C.Ho}\cite{I.Siomina} to describe the coupling relations among power, load and user data rates with equations. Another critical step is to find the optimal operating power at RRHs and the BBU pool in an efficient manner. We focus on iterative algorithms which can be considered as efficient algorithms for optimization problems with multiple coupling variables. The basic idea behind iterative algorithms is to decompose a multiple-variable optimization problem into several single-variable optimization problems by fixing some variables and then solve these single-variable optimization problems iteratively to find the optimal solution. To our best knowledge, we are the first to study the joint power optimization problem in C-RAN. Our main contributions can be described as follows:

1) Based on the load coupling model, we formulate the joint power optimization problem subject to QoS requirements from users in terms of data rates, considering both wireless transmission power at RRHs and baseband processing power at the BBU pool. The formulation theoretically characterize the tradeoff between these two kinds of power consumption.

2) We prove that operating at full load shall not be optimal in minimizing total power consumption in C-RAN. This result is different from that in the traditional BS system where only wireless transmission power is considered.

3) We design an efficient iterative algorithm to solve the problem of joint power optimization in C-RAN. In the BBU pool, the number of active BBUs reflects load of RRHs. At the first step, we concentrate on solving the joint power optimization problem with a fixed number of active BBUs. Then, we attempt to find the optimal power solution by iteratively changing the number of active BBUs.

We also perform extensive simulations to validate the theoretical and algorithmic work. The results show the proposed algorithm achieves reduced power consumption compared to existing schemes.

The rest of the paper is organized as follows. We give a brief overview on related work in Section II. In Section III, we describe the system model and formulate the joint power optimization problem in C-RAN. In Section IV, we design an iterative algorithm to find the optimal operating power at RRHs and the BBU pool. Simulation results are presented and analyzed in Section V. Finally, we conclude this paper with a summary in Section VI.

\section{RELATED WORKS}
To minimize energy consumption in cellular networks, much work has been done on power optimization in RANs. Unlike the traditional BS system, C-RAN involves three kinds of power, i.e., wireless transmission power, baseband processing power and backhaul transmission power. However, most of existing work only considers one of these three kinds of power.

In the traditional BS system, optimization of wireless transmission power under different kinds of QoS constraints is particularly concerned \cite{G.Bacci,M.Peng_2,C.W.Tan,L.Musavian,F.Sohrabi,C.Xiong,Q.Xu}, which is also beneficial to alleviate interference among cells. Under rate constraints, \cite{G.Bacci}\cite{M.Peng_2} attempt to minimize the wireless transmission power consumption in heterogeneous networks. In \cite{C.W.Tan,L.Musavian,F.Sohrabi}, the wireless transmission power consumption minimization problem is investigated with consideration of outage probability constraints. In order to improve the energy efficiency, power optimization is jointly performed with subcarrier allocation in OFDM networks in \cite{C.Xiong}\cite{Q.Xu}. To tackle the power optimization problem more practically, the authors in \cite{C.Ho} employ an analytical Signal to Interference and Noise Ratio (SINR) model that takes into account the load of each cell, resulting in a non-linear load coupling equation. This load coupling equation has also been shown to give a good approximation for more complicated load models in cellular networks that capture the dynamic nature of arrivals and service periods of data flows in the network. They use the load coupling equation to optimization wireless transmission power at BSs, and also prove that operating at full load at each BS is optimal in minimizing the wireless transmission power consumption.

There also exists some work on optimizing baseband processing power in C-RANs. By virtualization, the baseband processing resources in the BBU pool are dynamically shared among all RRHs. In \cite{G.Zhai}, a dynamic programming scheme is proposed to minimize the baseband processing power consumption in the case of dynamic cell traffic load. However, the computational complexity of the proposed scheme is exponential. In order to reduce the computational complexity, the authors in \cite{M.Qian} propose a BBU virtualization scheme that minimizes the baseband processing power consumption with a linear computational complexity order.

\section{SYSTEM MODEL}
\begin{figure}[t]
  \centering
  \includegraphics[width=0.35\textwidth]{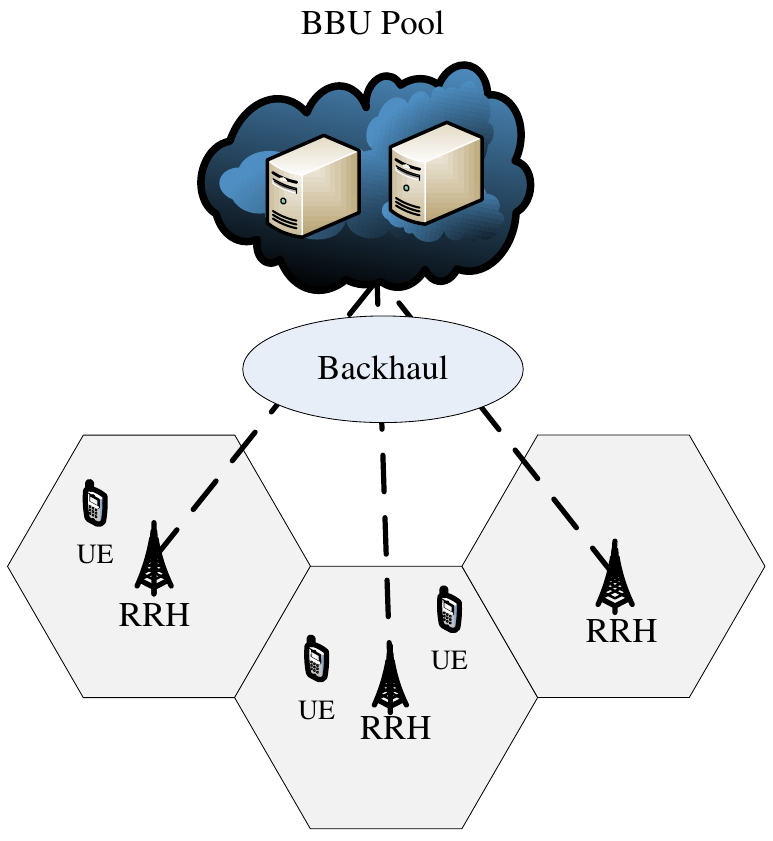}
  \caption{C-RAN architecture}\label{C-RAN}
\end{figure}

As shown in Fig. \ref{C-RAN}, we consider a C-RAN based cellular network, which consists of $ N $ RRHs and $ M $ BBUs. We denote the set of all RRHs by $\mathcal{N} = \{1,2,3,...,N\}$. Let $\mathcal{J}=\{1,2,...,J\}$ and $\mathcal{J}_{i}$ denote the set of users in the C-RAN and the set of users served by RRH $i$, respectively.

Without loss of generality, we assume that QoS requirements from users are expressed in terms of data rates. Let $r_{j}$ denote the data rate demand of user $j$ ($j\in \mathcal{J}$). The data rate demand vector is denoted by $\bm{r} = (r_{1}, ..., r_{J})^{T}$.

In this paper, boldface lower case, lower case and calligraphic symbols represent vectors, scalars and sets, respectively.

\subsection{Load Coupling Model}
In this paper, load is defined as the fractional usage of time-frequency resources at RRH. In LTE networks, time-frequency resources are allocated in the form of RBs. Let $\bm{x}$ denote the load vector, we have $ \bm{x} = (x_{1}, ..., x_{N})^{T} $, where $ x_{i} $ represents load of RRH $i$ ($i \in \mathcal{N}$). Given data rate demand vector $\bm{r}$, load couples with wireless transmission power at RRHs. We assume wireless transmission power of RRH $i$ is $p_{i}$. The power vector is denoted by $\bm{p} = (p_{1}, ..., p_{N})^{T}$.

Let $ x_{ij} $ denote the fraction of RBs allocated to user $ j $ by RRH $ i $. We have
 \begin{equation*}
	x_{ij} = r_{j} / R_{ij},
  \end{equation*}
where $ R_{ij} $ is the achievable data rate for user $j$ when it is associated with RRH $i$. It should be noted that $ x_{ij} $ is a real variable. To obtain  $ R_{ij} $, we model the SINR of user $j$ associated with RRH $i$ as follows:
  \begin{equation}\label{sinr}
  \gamma_{ij} = \frac{p_{i}h_{ij}}{\sum_{k \in \bm{N} \setminus \{ i \} } p_{k}x_{k}h_{kj} + \sigma^{2}},
  \end{equation}
where $ h_{ij} $ is the channel gain between RRH $ i $ and user $ j $, and $ \sigma^{2} $ represents average noise power. Then, based on the Shannon capacity theorem, $R_{ij}$ can be expressed as follows.
  \begin{equation*}
  R_{ij} = W log_{2}(1 + \gamma_{ij}),
  \end{equation*}
where $ W $ is the system bandwidth.

Load of RRH $ i $ can be obtained by summing the resource usage fractions over all user in $ \mathcal{J}_{i} $, which is represented as follows:
  \begin{equation}\label{load}
  x_{i} = \sum_{j \in \mathcal{J}_{i}} x_{ij} = \sum\limits_{j \in \mathcal{J}_{i}} \frac{r_{j}} {W log_{2}(1 + \gamma_{ij})},
  \end{equation}
where $ 0 \leq x_{i} \leq 1 $. Intuitively, $ x_{i} $ can be interpreted as the transmission probability of RRH $i$ on all RBs.

To make the notation more compact, the following Non-Linear load Coupling Equation (NLCE) is introduced from \cite{C.Ho}.
  \begin{equation}\label{nlce}
   \bm{x} = \bm{f}(\bm{r},\bm{x},\bm{p}),
  \end{equation}
where $ \bm{f} = (f_{1}, ..., f_{N})^{T} $.

In the NLCE, load vector $ \bm{x} $ appears in both sides of the equation and cannot be simply expressed as a fixed-point solution in the closed form. The reason is that load at different RRHs couples with each other. That is to say, load $ x_{i} $ of RRH $ i $ affects load $ x_{k} $ of RRH $ k $ ($ \neq i $), which would then in turn affect load $x_{i}$.

\subsection{Power Consumption Model}

  Regarding the total power consumption in C-RAN where power consumed by RRHs as well as that by the BBU pool are concerned. It can be expressed as follows \cite{D.Sabella}:
  \begin{equation*}
  P_{total} = \sum\limits_{i \in \mathcal{N}}P_{R,i} + P_{B},
  \end{equation*}
where $P_{R,i}$ and $P_{B}$ denote the power consumption at RRH $i$ and that at the BBU pool, respectively.

  As for the power consumption at RRH $i$, $P_{R,i}$ can be expressed as:
  \begin{equation*}
  P_{R,i} = \frac{x_{i} \cdot p_{i}}{\eta} + P_{RF},
  \end{equation*}
where $ \eta $ denotes the power amplifier efficiency, and $ P_{RF} $ denotes the circuit power consumption for a RRH.

  The baseband processing power consumption at the BBU pool is linear with the number of active BBUs as well as the average utilization of each active BBU. Thus, $P_{B}$ can be expressed as follows:
  \begin{equation}\label{P_B}
  P_{B} = m \cdot (P_{0} + \Delta_{p} \cdot P_{max} \cdot y ),
  \end{equation}
where $ m $ denotes the number of active BBUs, $ P_{0} $ and $ P_{max} $ are the power consumption of BBU in idle mode and in full usage mode, respectively. Besides, $ \Delta_{p} $ is the slope of the equivalent linear power model which depends on the specific server adopted, and $ y $ denotes the average utilization of each active BBU.

  Furthermore, the number of active BBUs and the average utilization of each active BBU are related to load of each RRH. Based on this fact, parameters $m$ and $y$ can be expressed as follows:
  \begin{equation}\label{number_of_BBUs}
  m = \left\lceil \frac{\sum_{i \in \mathcal{N}} k \cdot x_{i}}{X_{cap}} \right\rceil,
  \end{equation}
  \begin{equation}\label{average_utilization}
  y = \frac{\sum_{i \in \mathcal{N}} k \cdot x_{i}}{m \cdot X_{cap}},
  \end{equation}
where $ k $ is a relevant coefficient \cite{D.Sabella}, $ X_{\text{cap}} $ denotes the baseband processing capacity of one BBU in the BBU pool and the notation $ \lceil\cdot  \rceil $ is the ceiling function. In addition, load balancing is considered in calculating the average utilization of each active BBU according to Eq. (\ref{number_of_BBUs}), which is the same as that applying in \cite{D.Sabella}.

  Substituting Eqs. (\ref{number_of_BBUs}) and (\ref{average_utilization}) into Eq. (\ref{P_B}) , $P_B$ can be further expressed as follows:
  \begin{equation*}
  P_{\text{B}} = \left\lceil \frac{\sum_{i \in \mathcal{N}} k \cdot x_{i}}{X_{cap}} \right\rceil P_{0} + \frac{\Delta_{p} \cdot P_{max}}{X_{cap}}\sum\limits_{i \in \mathcal{N}}k \cdot x_{i}.
  \end{equation*}

  In summary, the total power consumption in C-RAN, where both wireless transmission power and baseband processing power are concerned, can be written as follows:
  \begin{equation}\label{P_total}
  P_{total} = m \cdot P_{0} + \sum\limits_{i \in \mathcal{N}}(a \cdot p_{i} \cdot x_{i} + b \cdot x_{i} + P_{RF}),
  \end{equation}
  where $ a = 1/{\eta} $, $ b = \Delta_{p} \cdot P_{max} \cdot k/X_{cap} $.

\subsection{Problem Formulation}

  In this paper, we consider the power consumption minimization problem in C-RAN while guaranteeing the QoS requirements of users. As backhaul transmission power is mainly determined by given user data rates, only wireless transmission power at RRHs and baseband processing power at the BBU pool are concerned. Mathematically, the power optimization problem is formulated as follows.
  \begin{subequations}\label{P0}
  \begin{align}
  \mathcal{P}0: \ \  &\underset{\bm{p}}{\text{minimize}}
  & & P_{total}\\
  & \text{subject to}
  & &\bm{x} = \bm{f}(\bm{r},\bm{x},\bm{p}),\\
  &&&\left\lceil \frac{\sum_{i \in \mathcal{N}} k \cdot x_{i}}{X_{cap}} \right\rceil \le m,\\
  &&&0 \le x_i \le 1, \quad \forall i \in \mathcal{N},\\
  &&&p_{min} \le p_{i} \le p_{max}, \quad \forall i \in \mathcal{N}.
 \end{align}
 \end{subequations}
 The objective is total power consumption in C-RAN, which is expressed in Eq. (\ref{P_total}). Constraint (\ref{P0}c) means that the number of active BBUs should be sufficient to support the requirements on the baseband processing capacity. Constraint (\ref{P0}d) means that load of each RRH should be between $0$ and $1$. In constraint (\ref{P0}e), $ p_{\text{min}} $ and $ p_{\text{max}} $ are the lower and upper bounds of each RRH's wireless transmission power, respectively.

\section{JOINT POWER OPTIMIZATION IN C-RAN}
In the traditional BS system, it has been proved that minimum power consumption can be achieved by operating at full load at each BS \cite{C.Ho}. However, in this section, we prove that such proposition cannot be applied to C-RAN, where both wireless transmission power and baseband processing power should be concerned. Due to the coupling relations between power consumption at RRHs and that at the BBU pool, problem $\mathcal{P}0$ is complicated to solve directly. As $\mathcal{P}0$ is a multiple-variable optimization problem, we attempt to solve it by an efficient iterative algorithm. The basic idea is to decompose problem $\mathcal{P}0$ into several single-variable optimization problems by fixing some variables and then solve these single-variable optimization problems iteratively to find the optimal solution.

\subsection{Analysis of Optimal Solution of $\mathcal{P}0$}
In the traditional BS system, it has been proved that the minimal wireless transmission power consumption can be achieved by operating at full load at each BS. Thus, we have the following proposition.

 \begin{proposition}
 In C-RAN, when only wireless transmission power consumption at RRHs is considered for problem $\mathcal{P}0$, optimal load of RRH $ i $ satisfies $ x_{i} = 1, \forall i \in \mathcal{N} $.
 \end{proposition}
 \emph{Proof:} The proof is similar to that of \emph{Theorem 1} in \cite{C.Ho}.

 \qed

However, operating at full load at each RRH may not be optimal in minimizing the total power consumption in C-RAN. In order to prove this conclusion, we firstly have the following lemma when baseband processing power of the BBU pool is considered.

 \begin{lemma}
 In C-RAN, when all RRHs fix their wireless transmission power except RRH $i$, the total power consumption of the network may not always increase with the increase of wireless transmission power of RRH $i$.
 \end{lemma}
 \emph{Proof:} The power consumption at RRH $i$ can be expressed as follows:
 \begin{equation*}
 \begin{aligned}
  P_{R,i} &= a \cdot p_{i} \cdot x_{i} + b \cdot x_{i} + P_{RF}\\
  & = \sum_{j \in \mathcal{J}_i} \frac{r_{j} (a \cdot p_{i} + b)}{W log_{2}(1 + c_{ij} \cdot p_{i})} + P_{RF},
  \end{aligned}
 \end{equation*}
 where $ c_{ij} = \frac{h_{ij}} {{\sum_{k \in \mathcal{N} \setminus \{ i \} } p_{k}h_{kj}x_{k} + \sigma^{2}}} $. It can be seen that $P_{R,i}$ is a function of $p_i$. Thus, we take the partial of $P_{R,i}$ with respect to $p_{i}$ and obtain:
 \begin{equation}\label{partial}
  \frac{\partial P_{R,i}}{\partial p_{i}}
  = \sum\limits_{j \in \mathcal{J}_i} \frac{r_{j}}{W} \cdot \frac{a(1 + c_{ij}p_{i})log_{2}(1 + c_{ij}p_{i}) - c_{ij}(ap_{i} + b)} {(log_{2}(1 + c_{ij}p_{i}))^2(1 + c_{ij}p_{i})}.
 \end{equation}
 The denominator of Eq. (\ref{partial}) is positive when $ p_i > 0 $. Besides, the numerator of Eq. (\ref{partial}) is a monotone increasing function of $ p_i $ when $ p_i > 0 $.
 Then, we have $ \frac{\partial g_{i}}{\partial p_{i}} < 0 $ when $ p_{i} \to 0 $, and $ \frac{\partial g_{i}}{\partial p_{i}} > 0 $ when $ p_{i} \to \infty $. That is to say, when $p_i$ increases, the power consumption at RRH $i$ may not always increase. Therefore, the total power consumption of the network may not always increase with the increase of wireless transmission power of RRH $i$. This completes the proof.

 \qed

Based on \emph{Lemma 1}, we can obtain the following proposition, which illustrates a property of the optimal solution of problem $\mathcal{P}0$.
 \begin{proposition}
 In C-RAN, when wireless transmission power consumption at RRHs and baseband processing power consumption at the BBU pool are jointly considered for problem $\mathcal{P}0$, optimal load of RRH $ i $ would not satisfy $ x_{i} = 1, \forall i \in \mathcal{N} $.
 \end{proposition}
 \emph{Proof:} Since the rate demand of each user is positive, the load for an arbitrary RRH $i$ satisfies $ x_{i} > 0 $. Assume in the optimal solution, the load of RRH $i$ is $ x_{i}^{\star} = 1 $ and wireless transmission power of RRH $i$ is $ p_{i}^{\star} $. Keep wireless transmission power and load of all other RRHs except that of RRH $i$ fixed. We consider another power-load pair $ (p', x') $ for RRH $ i $. The new wireless transmission power is $ p' = p_{i}^{\star} + \epsilon, \epsilon > 0 $. According to Eqs. (\ref{sinr}) and (\ref{load}), the new load is $ x' = x_{i}^{\star} - \epsilon', \epsilon' > 0 $.

 We can choose a small $ \epsilon > 0 $ such that the change of the whole network load is negligible. Therefore, the number of active BBUs is supposed to keep fixed. According to \emph{Lemma 1}, we claim that the total network power consumption may not be always increased when the wireless transmission power of RRH $ i $ increases. In other words, optimal load of RRH $ i $ would not satisfy $ x_{i} = 1, \forall i \in \mathcal{N} $. This completes the proof.

  \qed

Although \emph{Proposition 2} illustrates a property of optimal solution of problem $\mathcal{P}0$, such problem is complicated to solve directly. Solving problem $\mathcal{P}0$ faces two difficulties. One is that the ceiling function in constraint (\ref{P0}c) makes the problem intractable to be solved. The other is the coupling relations between power consumption at RRHs and that at the BBU pool.

To tackle the two difficulties, we firstly fix the number of active BBUs and ignore constraint (\ref{P0}c). Secondly, an efficient iterative algorithm is proposed to solve the optimization problem with multiple coupling variables. Finally, the optimal solution can be found through iterating the number of active BBUs. The detailed procedures will be discussed in the following two subsections.

\subsection{Solution of $\mathcal{P}0$ under Given Active BBUs' Number}
Since the ceiling function in constraint (\ref{P0}c) makes problem $\mathcal{P}0$ intractable to be solved, we can first fix the number of active BBUs as $m$ and convert problem $\mathcal{P}0$ to:
 \begin{subequations}\label{P1}
  \begin{align}
  \mathcal{P}1: \ \  &\underset{\bm{p}}{\text{minimize}}
  & & \sum\limits_{i \in \mathcal{N}}(a \cdot p_{i} + b )\cdot x_{i} \\
  & \text{subject to}
  & &\bm{x} = \bm{f}(\bm{r},\bm{x},\bm{p}),\\
  &&&\sum_{i \in \mathcal{N}} k \cdot x_{i} \le m \cdot X_{cap},\\
  &&&0 \le x_i \le 1, \quad \forall i \in \mathcal{N},\\
  &&&p_{min} \le p_{i} \le p_{max},  \quad \forall i \in \mathcal{N}.
 \end{align}
 \end{subequations}
 It should be noted that constraint (\ref{P0}c) in problem $\mathcal{P}0$ has been converted to constraint (\ref{P1}c) in problem $\mathcal{P}1$.

 Denote $ s_{i} = p_{i}x_{i}, \forall i \in \mathcal{N} $, and $ \bm{s} = (s_{1}, ..., s_{N}) $. Eq. (\ref{load}) can be reformulated as
 \begin{equation*}
  s_{i} = \sum\limits_{j \in \mathcal{J}_i} \frac{r_{j}p_{i}} {W \text{log}_{2}(1 + \frac{p_{i}h_{ij}}{\sum_{k \in \mathcal{N} \setminus \{ i \} } s_{k}h_{kj} + \sigma^{2}})}.
  \end{equation*}

 By doing so, we can use the following compact notation instead of NLCE in Eq. (\ref{nlce}).
 \begin{equation*}
  \bm{s} = \bm{t}(\bm{s},\bm{p}),
  \end{equation*}
  where $ \bm{t} = (t_{1}, ..., t_{N})^{T} $.

 Then, we use parameter $ \bm{s} $ instead of parameter $ \bm{x} $ in problem $\mathcal{P}1$, and problem $\mathcal{P}1$ can be converted to:
 \begin{equation*}
  \begin{aligned}
  \mathcal{P}2: \ \  &\underset{\bm{p},\bm{s}}{\text{minimize}}
  & & \sum\limits_{i \in \mathcal{N}}(a + \frac{b}{p_i})\cdot s_i\\
  & \text{subject to}
  & &\sum_{i \in \mathcal{N}} \frac{k \cdot s_{i}}{p_{i}} \le m \cdot X_{\text{cap}},\\
  &&&t_i(\bm{s},\bm{p}) \le s_i, \quad \forall i \in \mathcal{N},\\
  &&&0 \le s_i \le p_i, \quad \forall i \in \mathcal{N},\\
  &&&p_{min} \le p_{i} \le p_{max}, \quad \forall i \in \mathcal{N}.
 \end{aligned}
 \end{equation*}


Due to the coupling relations between $ \bm{s} $ and $ \bm{p} $, and interactions among elements in $ \bm{s} $, it is difficult to determine whether problem $\mathcal{P}2$ is convex or not. Thus, problem $\mathcal{P}2$ is still difficult to be solved directly. To tackle this problem, we focus on iterative algorithms which are considered as a kind of efficient algorithms for optimization problems with multiple coupling variables. Firstly, we optimize $\bm{s}$ under fixed $\bm{p}$. Then, we optimize $ \bm{p} $ under fixed $ \bm{s} $ which is obtained in the first step. These two steps are iterated until convergence. Similar iterative algorithms have been adopted in \cite{D.T.Ngo}\cite{K.Shen} which have verified that the global sub-optimal solutions can be achieved.

In the following content, the above two steps in the proposed iterative algorithm will be discussed in detail.

 \subsubsection{Optimizing $ \bm{s} $ under fixed $ \bm{p} $}

 Denote $ \bm{p}^{(k)} $ as the value of $ \bm{p} $ in the $k$th iteration. Given $ \bm{p}^{(k)} $, problem $\mathcal{P}2$ can be expressed as:
 \begin{equation*}
  \begin{aligned}
  \mathcal{P}3: \ \  &\underset{\bm{s}}{\text{minimize}}
  & & \sum\limits_{i \in \mathcal{N}}(a + \frac{b}{p_i^{(k)}})\cdot s_i\\
  & \text{subject to}
  & &\sum_{i \in \mathcal{N}} \frac{k \cdot s_{i}}{p_{i}^{(k)}} \le m \cdot X_{cap},\\
  &&&0 \le s_i \le p_i^{(k)}, \quad \forall i \in \mathcal{N},\\
  &&&t_i(\bm{s},\bm{p}^{(k)}) \le s_i, \quad \forall i \in \mathcal{N}.
 \end{aligned}
 \end{equation*}

 It has been proved in \cite{I.Siomina} that the sufficient condition of the solution existence of problem $P3$ is that the following linear programming problem has an optimal solution.
 \begin{subequations}\label{P4}
  \begin{align}
  \mathcal{P}4: \ \  &\underset{\bm{s}}{\text{minimize}}
  & & \sum\limits_{i \in \mathcal{N}}(a + \frac{b}{p_i^{(k)}})\cdot s_i\\
  & \text{subject to}
  & &\sum_{i \in \mathcal{N}} \frac{k \cdot s_{i}}{p_{i}^{(k)}} \le m \cdot X_{cap},\\
  &&&0 \le s_i \le p_i^{(k)}, \quad \forall i \in \mathcal{N},\\
  &&&\bm{H}^{(0)}\bm{s} + \bm{t}(\bm{s}^{(0)},\bm{p}^{(k)}) \le \bm{s}.
 \end{align}
 \end{subequations}
 Let $\bm{\rho} =(\rho_1,...,\rho_N)  $ and $\bm{\rho} = \bm{H}^{(0)}\bm{s} + \bm{t}(\bm{s}^{(0)},\bm{p}^{(k)})$, where $\bm{s}^{(0)} = (0, ..., 0)^{T} $ and $ \bm{H} = (h_{ik})_{N \times N}$. As for an arbitrary element $h_{ik}, \forall i, k \in \mathcal{N}$ in $ \bm{H} $,
 \begin{equation*}
 h_{ik} =
 \left\{
 \begin{array}{c}
 ln(2)\sum_{j \in \mathcal{J}_i}\frac{d_{j}p_{k}h_{jk}}{Wh_{ij}} , \quad i \neq k ,\\
 0, \quad otherwise.
 \end{array}
 \right.
 \end{equation*}
 It should be emphasized that the expression in constraint (\ref{P4}d) is equivalent to $\rho_i \le s_i, \forall i\in \mathcal{N}$. Such expression is also applied in the following content.

 Besides, according to \emph{Theorem 8} in \cite{I.Siomina}, the necessary condition of the solution existence of problem $\mathcal{P}3$ is that $\mathcal{P}4$ exists a feasible solution, and such solution is a lower bound of the solution of problem $P3$. Let $ \bm{s}^{\star} $ and $ \overline{\bm{s}} $ denote the feasible solutions of problem $\mathcal{P}3$ and $\mathcal{P}4$, respectively. Then, $ \overline{\bm{s}} \le \bm{s}^{\star} $.
 Based on such fact, a tightly lower bound could only be found in the interval $ \bm{s} \ge \overline{\bm{s}} $. Hence, we can construct the following Gauss-Seidel update form
 \begin{equation}\label{G-S_form}
  \overline{\bm{s}}^{(l+1)} = (\bm{I} - \bm{H}^{(0)})^{-1}( \bm{t}(\overline{\bm{s}}^{(l)},\bm{p}^{(k)}) - \bm{H}^{(0)} \overline{\bm{s}}^{(l)} ).
 \end{equation}

Based on Eq. (\ref{G-S_form}), we have the following lemma, which points out the method of solving problem $\mathcal{P}3$.
 \begin{lemma}
 If problem $\mathcal{P}4$ is feasible and $ \bm{s}^{\star} \ge \overline{\bm{s}}^{(l)} $, the Gauss-Seidel update formulation in Eq. (\ref{G-S_form}) is convergent, and $ \overline{\bm{s}}^{(l)} \le \overline{\bm{s}}^{(l+1)} \le \bm{s}^{\star} $.
 \end{lemma}
 \emph{Proof:} Assuming that problem $\mathcal{P}4$ is feasible and $ \bm{s}^{\star} \ge \overline{\bm{s}}^{(l)} $. We obtain $ \bm{s}^{\star} \ge \bm{t}(\bm{s}^{\star},\bm{p}^{(k)}) \ge \bm{H}^{(0)}\bm{s}^{\star} + \bm{t}(\bm{s}^{\star},\bm{p}^{(k)}) \ge \bm{H}^{(0)}\overline{\bm{s}}^{(l)} + \bm{t}(\overline{\bm{s}}^{(l)},\bm{p}^{(k)}) $. Replace the constraint (\ref{P4}d) with new constraints $ \bm{H}^{(0)}\bm{s} + \bm{t}(\overline{\bm{s}}^{(l)},\bm{p}^{(k)}) \le \bm{s} $ and $ \overline{\bm{s}}^{(l)} \le \bm{s}$. It can be seen that $ \bm{s}^{\star} $ satisfies all the constraints and problem $\mathcal{P}4$ with new constraints is still feasible. Denote $ \overline{\bm{s}}^{(l+1)} $ the solution of problem $\mathcal{P}4$ with new constraints. We obtain $ \overline{\bm{s}}^{(l)} \le \overline{\bm{s}}^{(l+1)} \le \bm{s}^{\star} $. Since the series $\{..., \overline{\bm{s}}^{(l)},\overline{\bm{s}}^{(l+1)},... \}$ is monotonically nondecreasing and has an upper bound, such series is convergent. This completes the proof.

 \qed

  \begin{algorithm}[t]
  \caption{Optimizing $ \bm{s} $ Under Fixed $ \bm{p} $}             
  \label{DPC-Q11}                  
  \textbf{Initialize} $ \varepsilon $, $ \bm{p} $; let $ \overline{\bm{s}}^{(l)} = \bm{s}^{(0)} $ and $ l = 0 $;\\
  \If{matrix $ \bm{I} - \bm{H}^{(0)} $ is invertible}{
     $ \overline{\bm{s}}^{(l+1)} = (\bm{I} - \bm{H}^{(0)})^{-1}( \bm{t}(\overline{\bm{s}}^{(l)},\bm{p}^{(k)}) - \bm{H}^{(0)})$;\\
     \If{$ (\overline{\bm{s}}^{(l+1)} - \overline{\bm{s}}^{(l)})^T(\overline{\bm{s}}^{(l+1)} - \overline{\bm{s}}^{(l)}) > \varepsilon $ \& constraints (\ref{P4}b) and (\ref{P4}c) are satisfied}{
            $ l = l + 1 $;\\
            $ \overline{\bm{s}}^{(l+1)} = (\bm{I} - \bm{H}^{(0)})^{-1}( \bm{t}(\overline{\bm{s}}^{(l)},\bm{p}^{(k)}) - \bm{H}^{(0)})$;
     }
  }
  return $ \overline{\bm{s}}^{(l)}$;
  \end{algorithm}
 Based on \emph{Lemma 2}, we design an algorithm to optimize $ \bm{s} $ under fixed $ \bm{p} $, which is illustrated in Algorithm \ref{DPC-Q11}. According to \emph{Lemma 2}, we know the series $\{..., \overline{\bm{s}}^{(l)},\overline{\bm{s}}^{(l+1)},... \}$ is monotonically nondecreasing and satisfies $ \overline{\bm{s}}^{(l)} \le \overline{\bm{s}}^{(l+1)} $. Then if constraint (\ref{P4}b) or (\ref{P4}c) is not satisfied in the $l$th iteration, constraint (\ref{P4}b) or (\ref{P4}c) will not be satisfied in the following iterations. Hence, the iterative process will terminate if any of the above two constraint cannot be satisfied.

 \subsubsection{Optimizing $ \bm{p} $ under fixed $ \bm{s} $}

 Assume the result obtained from the above procedure is $ \bm{s}^{(k+1)} $. In this procedure, we will fix $ \bm{s}^{(k+1)} $, and try to solve problem $\mathcal{P}2$.

 Given $ \bm{s}^{(k+1)} $, problem $\mathcal{P}2$ can be expressed as
 \begin{subequations}\label{P5}
  \begin{align}
  \mathcal{P}5: \ \  &\underset{\bm{p}}{\text{minimize}}
  & & \sum\limits_{i \in \mathcal{N}}(a + \frac{b}{p_i})\cdot s_i^{(k+1)}\\
  & \text{subject to}
  & &\sum_{i \in \mathcal{N}} \frac{k \cdot s_{i}^{(k+1)}}{p_{i}} \le m \cdot X_{cap},\\
  &&&p_{\text{min}} \le p_{i} \le p_{\text{max}}, \quad \forall i \in \mathcal{N},\\
  &&&s_i^{(k+1)} \le p_i, \quad \forall i \in \mathcal{N},\\
  &&&t_i(\bm{s}^{(k+1)},\bm{p}) \le s_i^{(k+1)}, \quad \forall i \in \mathcal{N}.
 \end{align}
 \end{subequations}

As for problem $\mathcal{P}5$, the objective function is convex, and all the constraints are inequality constraints. Besides, all the constraint functions are convex. According to the definition of convex optimization problems in \cite{convex_optimization}, problem $\mathcal{P}5$ is a convex optimization problem, which can be effectively solved by convex optimization algorithms.
We first construct the Lagrange function for problem $\mathcal{P}5$, which can be expressed as follows:
 \begin{equation*}
 \begin{aligned}
 L(\bm{p},\bm{\lambda},\bm{\mu},\nu) &= \sum\limits_{i \in \mathcal{N}}(a + \frac{b}{p_i})\cdot s_i^{(k+1)}\\
  &+ \sum\limits_{i \in \mathcal{N}}\lambda_{i} \cdot (s_{i}^{(k+1)} - p_{i})\\
  &+ \sum\limits_{i \in \mathcal{N}}\mu_{i} \cdot (t_{i}(\bm{s}^{(k+1)},p_{i}) - s_{i}^{(k+1)})\\
  &+ \nu \cdot (\sum_{i \in \mathcal{N}} \frac{k \cdot s_{i}^{(k+1)}}{p_{i}} - m \cdot X_{\text{cap}}),
 \end{aligned}
 \end{equation*}
 where $ \nu $, $ \bm{\lambda} = \{\lambda_{i}, i \in \mathcal{N}\} $ and $ \bm{\mu} = \{\mu_{i}, i \in \mathcal{N}\} $ are the Lagrange multipliers associated with constraints (\ref{P5}b), (\ref{P5}d) and (\ref{P5}e), respectively. Therefore, the objective function of the dual problem of $\mathcal{P}5$ can be expressed as
 \begin{equation*}
  h(\bm{\lambda},\bm{\mu},\nu) = \mathop{\text{min}}\limits_{\bm{p} \ge \bm{0}} \ \ L(\bm{p},\bm{\lambda},\bm{\mu},\nu).
 \end{equation*}
 Then, the dual problem of $\mathcal{P}5$ is
 \begin{equation*}
  \mathop{\text{max}}\limits_{\bm{\lambda}\ge \bm{0},\bm{\mu} \ge \bm{0},\nu \ge 0} \ \ h(\bm{\lambda},\bm{\mu},\nu).
 \end{equation*}

 Since for a convex optimization problem, a strong duality exists \cite{convex_optimization}. In this case, the optimal solutions for the primal and dual problems are equal. That is to say, we can solve the dual problem instead of solving the primal problem $\mathcal{P}5$. The dual problem can be further expressed as
 \begin{equation}\label{dual_problem}
 \begin{aligned}
  h(\bm{\lambda},\bm{\mu},\nu) = &\sum\limits_{i \in \mathcal{N}} \ \mathop{\text{min}}\limits_{\bm{p} \ge \bm{0}} \  \{ a \cdot s_{i}^{(k+1)} + \frac{b \cdot s_{i}^{(k+1)}}{p_{i}} + \lambda_{i}(s_{i}^{(k+1)} - p_{i})\\
   &+ \mu_{i}(t_{i}(\bm{s}^{(k+1)},p_{i}) - s_{i}^{(k+1)}) + \nu \frac{k \cdot s_{i}^{(k+1)}}{p_{i}} \}.
  \end{aligned}
 \end{equation}

According to Eq. (\ref{dual_problem}), the dual problem can be solved by each RRH in a distributed manner. As for RRH $i$, it will try to solve its own minimization problem, which is expressed as follows:
 \begin{equation}\label{dual_problem2}
 \begin{aligned}
  &\mathop{\text{min}}\limits_{p_i \ge 0} \  \{ a \cdot s_{i}^{(k+1)} + \frac{b \cdot s_{i}^{(k+1)}}{p_{i}} + \lambda_{i} \cdot (s_{i}^{(k+1)} - p_{i}) \\
  &+ \mu_{i} \cdot (t_{i}(\bm{s}^{(k+1)},p_{i}) - s_{i}^{(k+1)}) + \nu \frac{k \cdot s_{i}^{(k+1)}}{p_{i}} \}.
  \end{aligned}
 \end{equation}

For fixed $\bm{\lambda}$, $\bm{\mu}$ and $\nu$, the optimal wireless transmission power of RRH $i$ can be calculated by each RRH $i$ through applying the Karush-Kuhn-Tucker (KKT) conditions \cite{convex_optimization} on Eq. (\ref{dual_problem2}), and we obtain
 \begin{equation}\label{KKT}
  - \frac{b \cdot s_{i}^{(k+1)}}{p_{i}^{2}} - \lambda_{i} + \mu_{i} \frac{\partial t_{i}(\bm{s}^{(k+1)},p_{i})}{\partial p_{i}} - \nu \frac{k \cdot s_{i}^{(k+1)}}{p_{i}^{2}} = 0.
 \end{equation}
 Then $ p_{i} $, which is the wireless transmission power of RRH $i$, can be calculated by solving Eq. (\ref{KKT}).

 The optimal values of the Lagrange multipliers that given the optimal wireless transmission power can be calculated by solving the dual problem of problem $\mathcal{P}5$. The dual problem can be further written as
 \begin{equation*}\label{dual_function3}
    \begin{aligned}
    &\sum\limits_{i \in \mathcal{N}} \mathop{\text{max}}\limits_{\bm{\lambda}\ge \bm{0}} \lambda_{i}(s_{i}^{(k+1)} - p_{i}) \\
    + &\sum\limits_{i \in \mathcal{N}} \mathop{\text{max}}\limits_{\bm{\mu}\ge \bm{0}} \mu_{i}(t_{i}(\bm{s}^{(k+1)},p_{i}) - s_{i}^{(k+1)}) \\
    + &\mathop{\text{max}}\limits_{\nu \ge 0} \nu (\sum_{i \in \mathcal{N}} \frac{ks_{i}^{(k+1)}}{p_{i}} - m \cdot X_{\text{cap}}),
    \end{aligned}
 \end{equation*}
 which is a differentiable function. Therefore, the value of the Lagrange multipliers $ \lambda_{i} $, $ \mu_{i} $ and $ \nu $ can be iteratively calculated through a gradient descent method. In each iteration, the Lagrange multipliers can be updated as follows:
 \begin{equation}\label{Lm_1}
  \lambda_{i}^{(l+1)} = [\lambda_{i}^{(l)} - \xi_{1}(s_{i}^{(k+1)} - p_{i}^{(l)})]^{+},
 \end{equation}
 \begin{equation}\label{Lm_2}
  \mu_{i}^{(l+1)} = [\mu_{i}^{(l)} - \xi_{2}(t_{i}(\bm{s}^{(k+1)},p_{i}^{(l)}) - s_{i}^{(k+1)})]^{+},
 \end{equation}
 \begin{equation}\label{Lm_3}
  \nu^{(l+1)} = [\nu^{(l)} - \xi_{3}(\sum_{i \in \mathcal{N}} \frac{k \cdot s_{i}^{(k+1)}}{p_{i}^{(l)}} - m \cdot X_{\text{cap}})]^{+},
 \end{equation}
 where $ l $ is the iteration index, $ \xi_{1} $, $ \xi_{2} $ and $ \xi_{3} $ are sufficiently small fixed step size for updating $ \lambda_{i} $, $ \mu_{i} $ and $ \nu $, respectively. The notation $[\cdot]^+$ is a projection on the positive orthant. There is a convergence guarantee for the optimal solution since the gradient of the problem (\ref{dual_problem}) satisfies the Lipchitz continuity condition \cite{convex_optimization}. 

 Based on the above analysis, we design an algorithm to optimize $ \bm{p} $ under fixed $ \bm{s} $, which is illustrated in Algorithm \ref{DPC-Q12}. In this algorithm, we assume $\xi_{1} = \xi_{2} = \xi_{3} = \xi$.
  \begin{algorithm}[t]
  \caption{Optimizing $ \bm{p} $ Under Fixed $ \bm{s} $}             
  \label{DPC-Q12}                  
  \textbf{Initialize} $\bm{p}^{0}$, $\bm{s}$, $ \lambda_{i}^{(0)}$, $ \mu_{i}^{(0)}, \forall i \in \mathcal{N} $, $ \nu^{(0)} $, $ \xi $, $ \varepsilon $; let $ l = 1 $;\\
  calculate $ p_{i}^{(l)} $ according to Eq. (\ref{KKT});\\
  update Lagrange multipliers according to Eqs. (\ref{Lm_1}), (\ref{Lm_2}) and (\ref{Lm_3});\\
  \If{$ (\bm{p}^{(l)} - \bm{p}^{(l-1)})^T(\bm{p}^{(l)} - \bm{p}^{(l-1)}) > \varepsilon $}{
        $l = l + 1$;\\
        calculate $ p_{i}^{(l)} $ according to Eq. (\ref{KKT});\\
        update Lagrange multipliers according to Eqs. (\ref{Lm_1}), (\ref{Lm_2}) and (\ref{Lm_3});\\
  }
  return $\bm{p}^{(l)}$.
  \end{algorithm}

 As discussed above, problem $\mathcal{P}2$ can be solved by iterating the two procedures which are presented in Algorithm \ref{DPC-Q11} and Algorithm \ref{DPC-Q12} in detail. Such iterative process is illustrated in an iterative power optimization algorithm, which is depicted in Algorithm \ref{DPC-Q1}.

  \begin{algorithm}[t]
  \caption{Iterative Power Optimization Algorithm}             
  \label{DPC-Q1}                  
  \textbf{Initialize}  $ \varepsilon $; let $ p_{i}^{(0)} = p_{max}, \forall i \in \bm{N} $, and $ k = 0 $;\\
  $ \bm{p} = \bm{p}^{(k)} $; apply Algorithm \ref{DPC-Q11} to solve problem $\mathcal{P}4$, and denote the result by $ \bm{s}^{(k+1)} $;\\
  $ \bm{s} = \bm{s}^{(k+1)} $; apply Algorithm \ref{DPC-Q12} to solve problem $\mathcal{P}5$, and denote the result by $ \bm{p}^{(k+1)} $;\\
  \If{ $ (\bm{p}^{(k+1)} - \bm{p}^{(k)})^T(\bm{p}^{(k+1)} - \bm{p}^{(k)}) > \varepsilon $}{
            $ k = k + 1 $;\\
            $ \bm{p} = \bm{p}^{(k)} $; apply Algorithm \ref{DPC-Q11} to solve problem $\mathcal{P}4$, and denote the result by $ \bm{s}^{(k+1)} $;\\
            $ \bm{s} = \bm{s}^{(k+1)} $; apply Algorithm \ref{DPC-Q12} to solve problem $\mathcal{P}5$, and denote the result by $ \bm{p}^{(k+1)} $;\\
  }
  $ \bm{p}^{\star} = \bm{p}^{(k+1)} $, $ x_{i}^{\star} = s_{i}^{(k+1)} / p_{i}^{(k+1)}, \forall i \in \mathcal{N} $;\\
  return $ \bm{p}^{\star}$, $ \bm{x}^{\star}$.
  \end{algorithm}


\subsection{Solution of $\mathcal{P}0$ under Varying Number of Active BBUs}

 In this subsection, we will attempt to find the optimal power solution of problem $\mathcal{P}0$ by iterating the number of active BBUs. First of all, keep all BBUs active and switch off BBUs one by one. Each time when a BBU is switched off, let the current number of active BBUs is $m$. We can obtain the optimal wireless transmission power of each RRH by applying Algorithm \ref{DPC-Q1}, and further obtain the current minimum total power consumption, which can be expressed as $P_{total}(m)$.

 Since the baseband processing capacity of active BBUs is enough at first, we can switch off some BBUs to reduce the baseband processing power consumption. However, when constraint (\ref{P0}c) in problem $\mathcal{P}0$ cannot be satisfied, BBUs cannot be further switched off. Then, the optimal total power consumption and the optimal number of active BBUs can be found through comparing the results under different numbers of active BBUs. The detailed procedures are described in Algorithm \ref{DPC-Q2}.

 \begin{algorithm}[t]
  \caption{Joint Power Optimization Algorithm}             
  \label{DPC-Q2}                  
  \textbf{Initialize} $ m = M $, $P^* = 0$ and $m^* = M$;\\
  calculate $ P_{total}(M) $ by applying Algorithm \ref{DPC-Q1};\\
  $P^* = P_{total}(M)$;\\
  $m = m - 1$;\\
  \While{ $ m > 0 $ \& constraint (8c) is satisfied}{
        calculate $ P_{total}(m) $ by applying Algorithm \ref{DPC-Q1};\\
        \If{$P^* < P_{total}(m)$}{
            $P^* = P_{total}(m)$;\\
            $m^* = m$;
        }
        $m = m - 1$;\\
  }
  return $ m^*$, $P^*$.
  \end{algorithm}

\section{PERFORMANCE EVALUATION}
In this section, numerical results are presented to validate our theoretical and algorithmic work on joint power optimization in C-RAN. To evaluate the performance of the proposed iteration based algorithm which attempts to search the optimal solution approximately for the joint power optimization problem, two benchmark algorithms as follows are involved for comparison.
 \begin{itemize}
  \item Exhaustive Search Algorithm (ESA): The optimal solution is found through an exhaustive search method, which searches all possible combinations of wireless transmission power and baseband processing power to obtain the minimal total power consumption in C-RAN.
  \item Transmission Power Optimization Algorithm (TPOA): TPOA is proposed in \cite{C.Ho}, where only wireless transmission power consumption is considered.
\end{itemize}

We can see that ESA can achieve the optimal solution. However, as an exhaustive search method, the highest computational complexity makes ESA impractical. Thus, we only use the results from ESA as the theoretical bounds with which the optimality of the proposed algorithm can be evaluated. On the other hand, with the results from TPOA, we can evaluate the advantages of joint power optimization performed by the proposed algorithm.

For the convenience of narration, we use `WTP' and `BPP' to denote wireless transmission power and baseband processing power in the following figures, respectively.

\subsection{Simulation Setup}
\begin{table}[h]
  \centering
  \caption{Simulation Parameters}
  \label{parameters}
  \begin{tabular}{cc}
  \toprule
  Parameters & Value\\
  \midrule
  Total Bandwidth of C-RAN & 10 MHz \\
  Inter-site Distance Between RRHs & 500 m \\
  Min Distance Between a RRH and a UE & 35 m \\
  Min RRH wireless transmission power & 12 dBm \\
  Max RRH wireless transmission power & 42 dBm \\
  Pathloss & $ 128.1 + 37.6 \text{log} 10 (d) $ \\
  Shadowing Deviation & 4 dB \\
  Noise Power Density & -174 dBm/Hz\\
  \bottomrule
  \end{tabular}
 \end{table}

 We consider a C-RAN based downlink cellular network, which consists of $12$ RRHs and $5$ BBUs at the most. Each RRH serves a unique group of users, which are uniformly distributed under the coverage of the RRH. According to the Third Generation Partnership Project (3GPP) LTE specification \cite{3GPP:Spec}, the bandwidth and time duration of a RB is 180 KHz and 0.5 ms, respectively. As for the power consumption model, we set $ \eta = 13.64\% $, $ P_{RF} = 12.8 $ dBm, $ P_{0} = 10 $ dBm, $ P_{max} = 100 $ dBm, $ \Delta_{p} = 0.44 $, $ X_{cap} = 314 $, $ k = 104.89 $. The above parameters are chosen according to \cite{D.Sabella}\cite{H.Holtkamp_2}. Other default simulation configurations are listed in Table \ref{parameters}, which are selected based on 3GPP LTE specification \cite{3GPP:Spec}. In the pathloss, $d$ is the distance between a user and its associated RRH in kilometers.


\subsection{Simulation Results}

 \begin{figure}[t]
  \centering
  \includegraphics[width=0.45\textwidth]{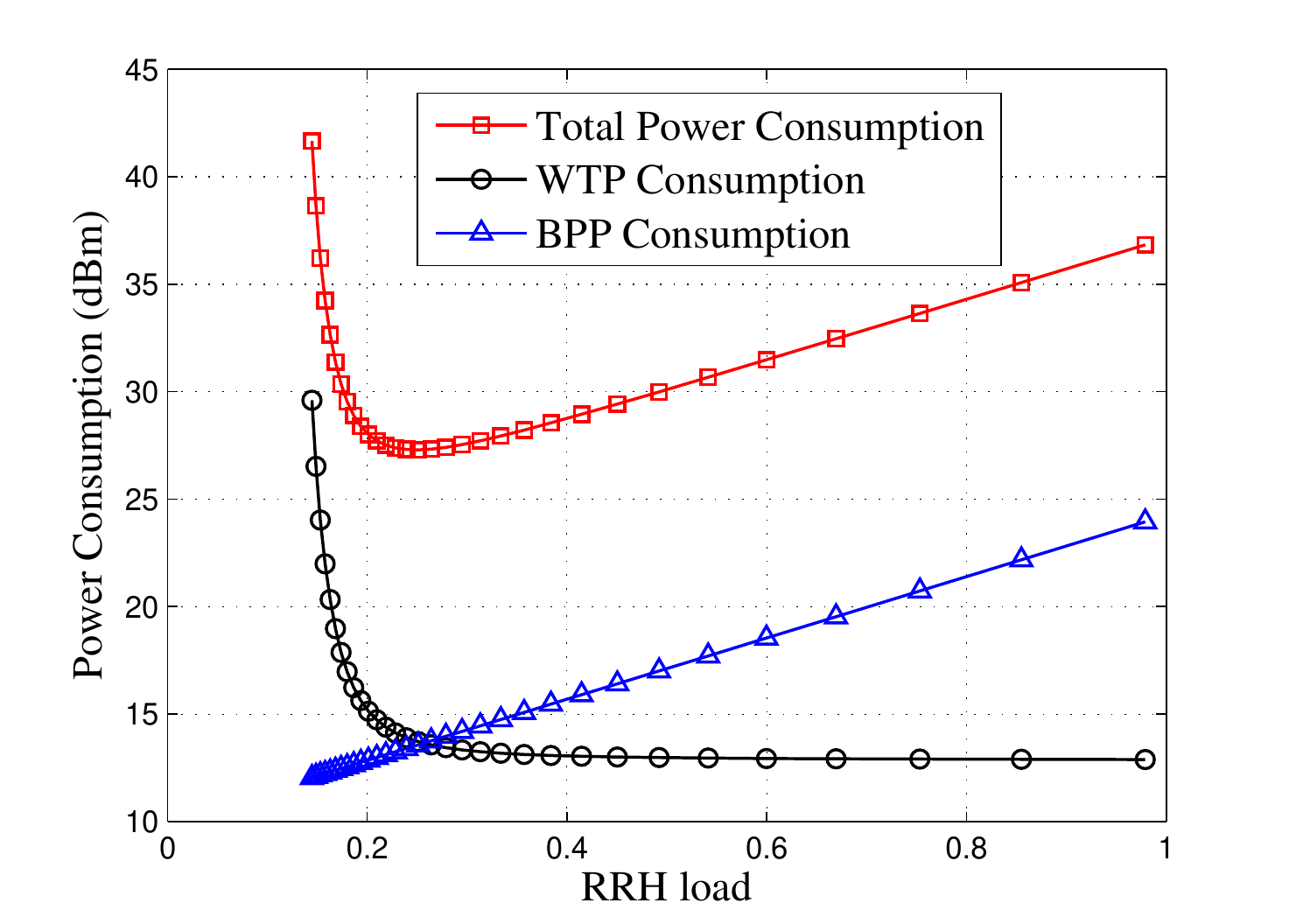}
  \caption{Power consumption vs. RRH load.}\label{fig1}
  \centering
  \includegraphics[width=0.45\textwidth]{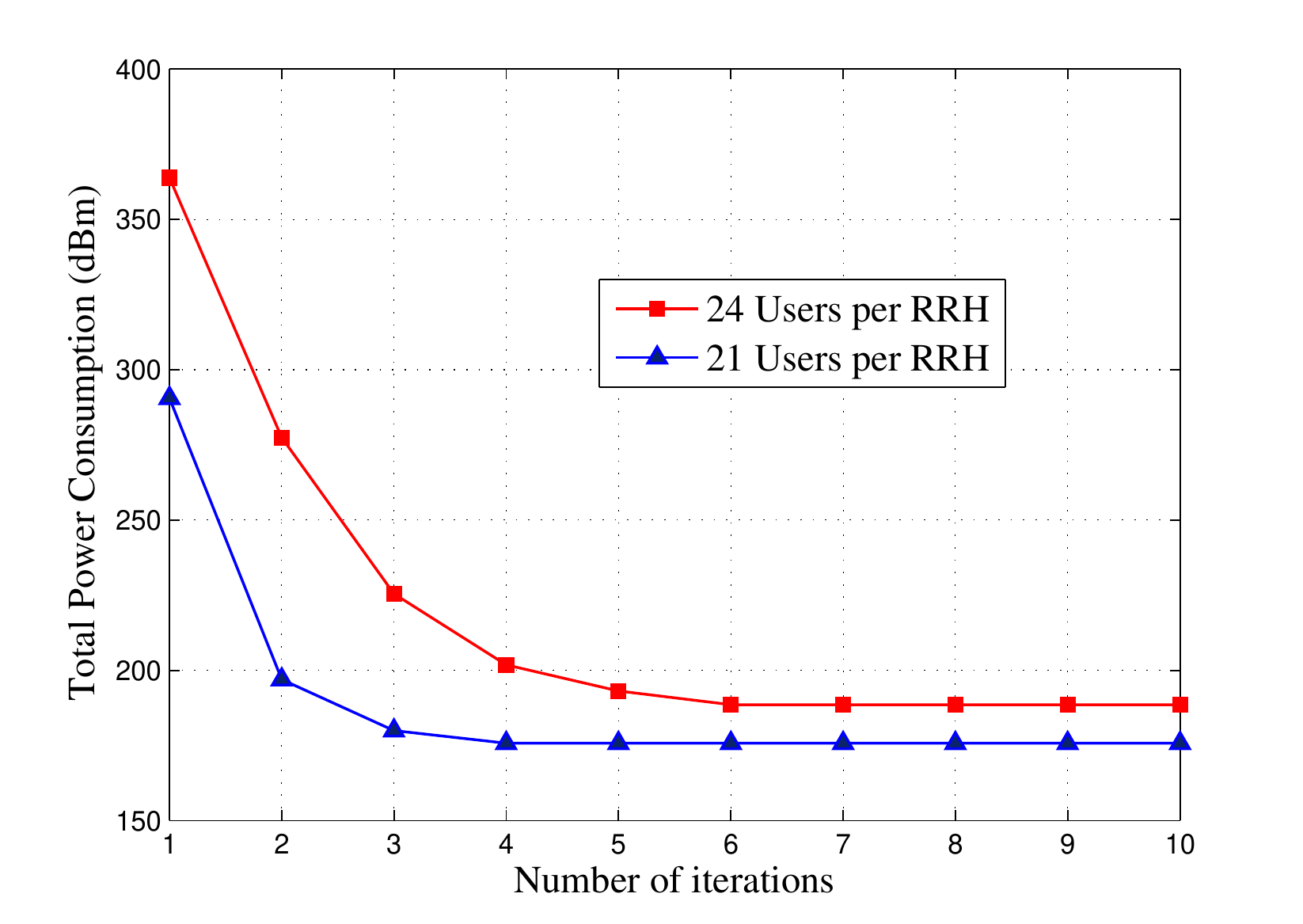}
  \caption{Total power consumption vs. Number of iterations.}\label{fig2}
 \end{figure}

In Fig. \ref{fig1}, we evaluate the total power consumption, the baseband processing power consumption and the wireless transmission power consumption in C-RAN when RRH load varies. In the simulation, $1$ BBU and $3$ RRHs are deployed in the network. Under the coverage of each RRH, $21$ users are uniformly distributed. The rate requirement of each user is set to $750$ Kbps. In the simulation, we vary the wireless transmission power of only one RRH, and fix the transmission power of the other two RRHs. Fig. \ref{fig1} illustrates that the total power minimization problem in C-RAN is convex, and there exists an optimal value for RRH load. Besides, according to Fig. \ref{fig1}, deploying full load can not achieve the optimal power consumption, which is in accordance with \emph{Proposition 2}. As expected, Fig. \ref{fig1} also shows that there is a trade-off between the wireless transmission power consumption and baseband processing power consumption. With the increasement of RRH load, the wireless transmission power consumption decreases while the baseband processing power consumption increases.

Fig. \ref{fig2} depicts the convergence of the proposed iterative power optimization algorithm under a given number of active BBUs. In the simulation, $3$ active BBUs and $6$ RRHs are deployed in the network. The rate requirement of each user is set to $1000$ Kbps, and $ \varepsilon $ adopted in the algorithm is set to be $ 10^{-2} $. According to Fig. \ref{fig2}, the proposed algorithm will converge in quite few iterations.



 \begin{figure}[t]
  \centering
  \includegraphics[width=0.45\textwidth]{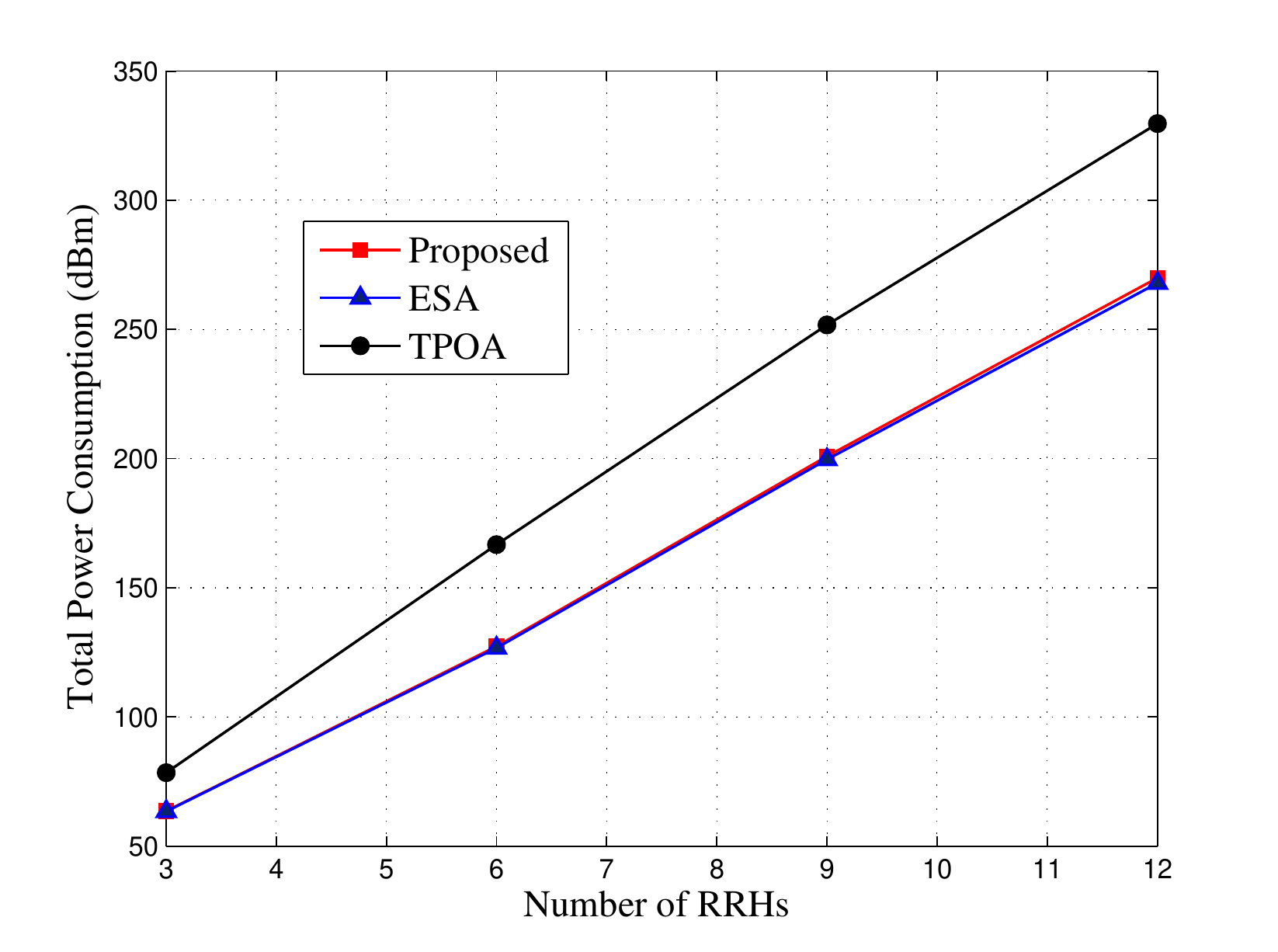}
  \caption{Total power consumption vs. number of RRHs.}\label{fig3}
  \centering
  \includegraphics[width=0.45\textwidth]{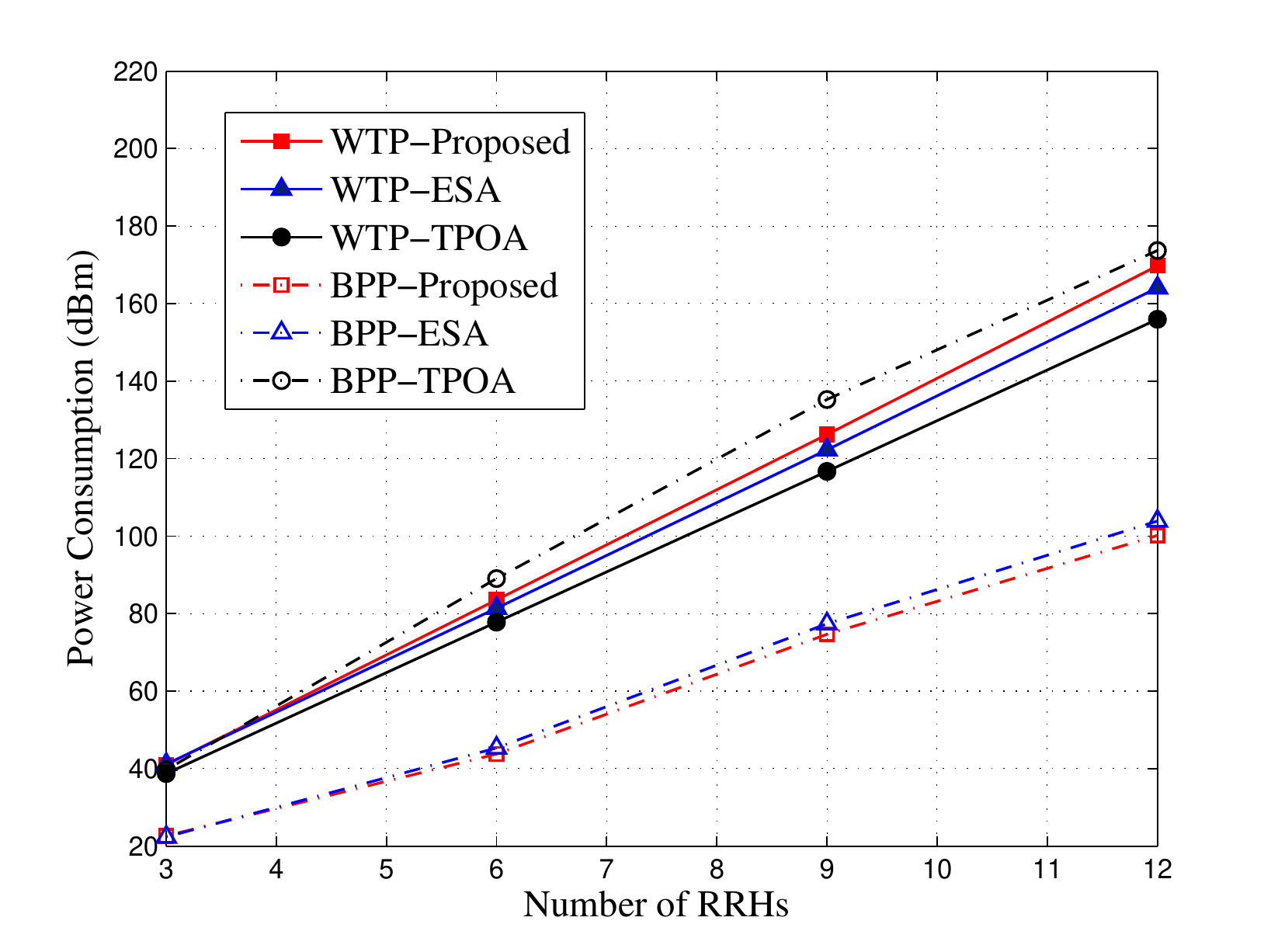}
  \caption{Power consumption vs. number of RRHs.}\label{fig4}
 \end{figure}

Fig. \ref{fig3} and Fig. \ref{fig4} show the power consumption varies with the number of RRHs deployed in C-RAN. In the simulations, $18$ users are deployed under the coverage of each RRH. The rate requirement of each user is set to $750$ Kbps. As expected, all kinds of power consumption, i.e., the total power consumption, the wireless transmission power consumption and the baseband processing power consumption, increase with the increasement of the number of RRHs. From Fig. \ref{fig3}, we can see that the proposed algorithm achieves similar performance as ESA where exhaustive search is applied and outperforms TPOA where joint power optimization is not considered. We can find the reason in Fig. \ref{fig4}. The total power consumption is the sum of the wireless transmission power consumption and baseband processing power consumption. In terms of wireless transmission power consumption, TPOA achieves the best performance. However, it consumes much more baseband processing power than the proposed algorithm and ESA. Thus, we can see that TPOA performs the worst when the total power consumption is considered.

 \begin{figure}[t]
  \centering
  \includegraphics[width=0.45\textwidth]{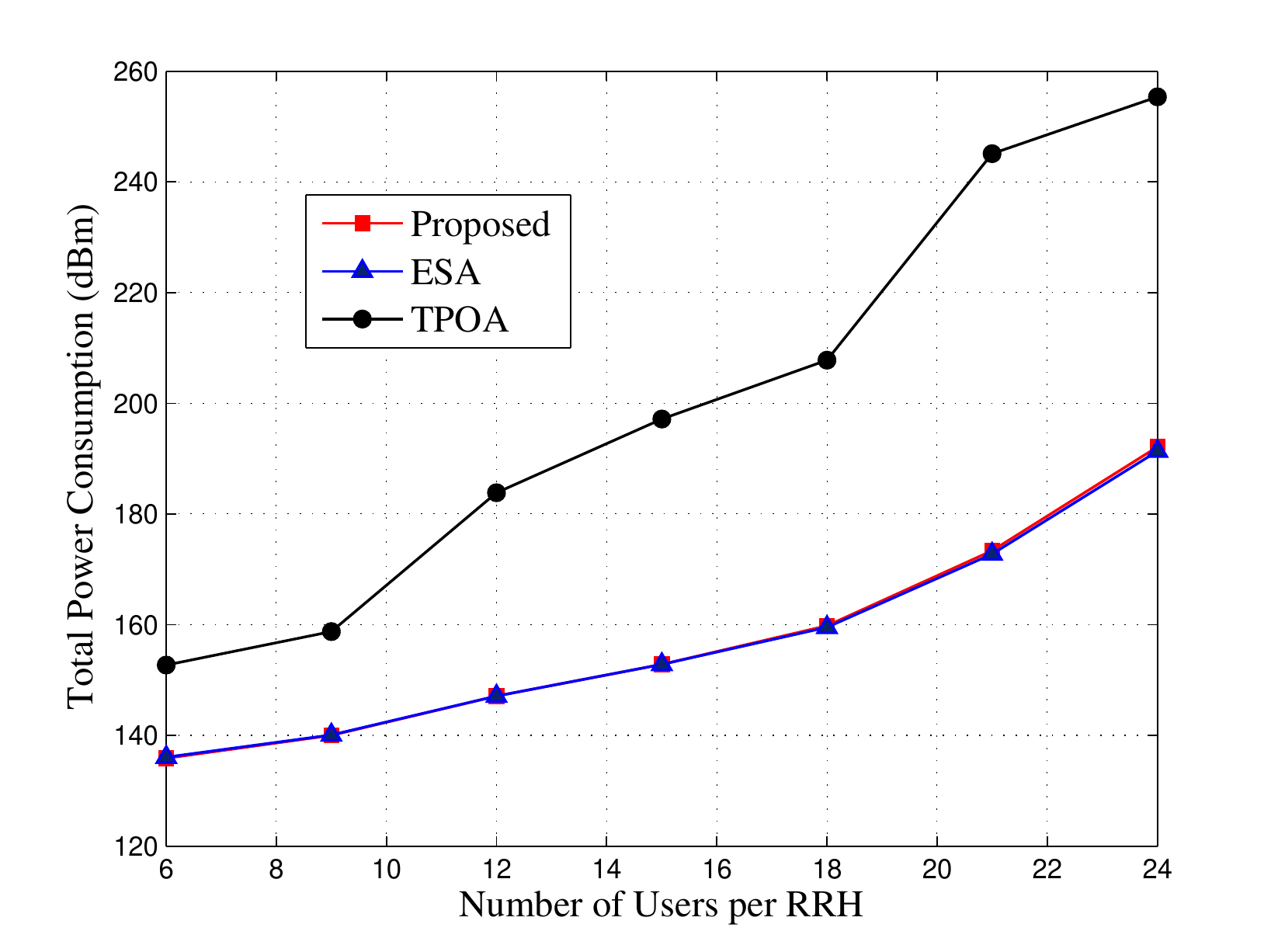}
  \caption{Total power consumption vs. Number of users per RRH.}\label{fig5}
  \centering
  \includegraphics[width=0.45\textwidth]{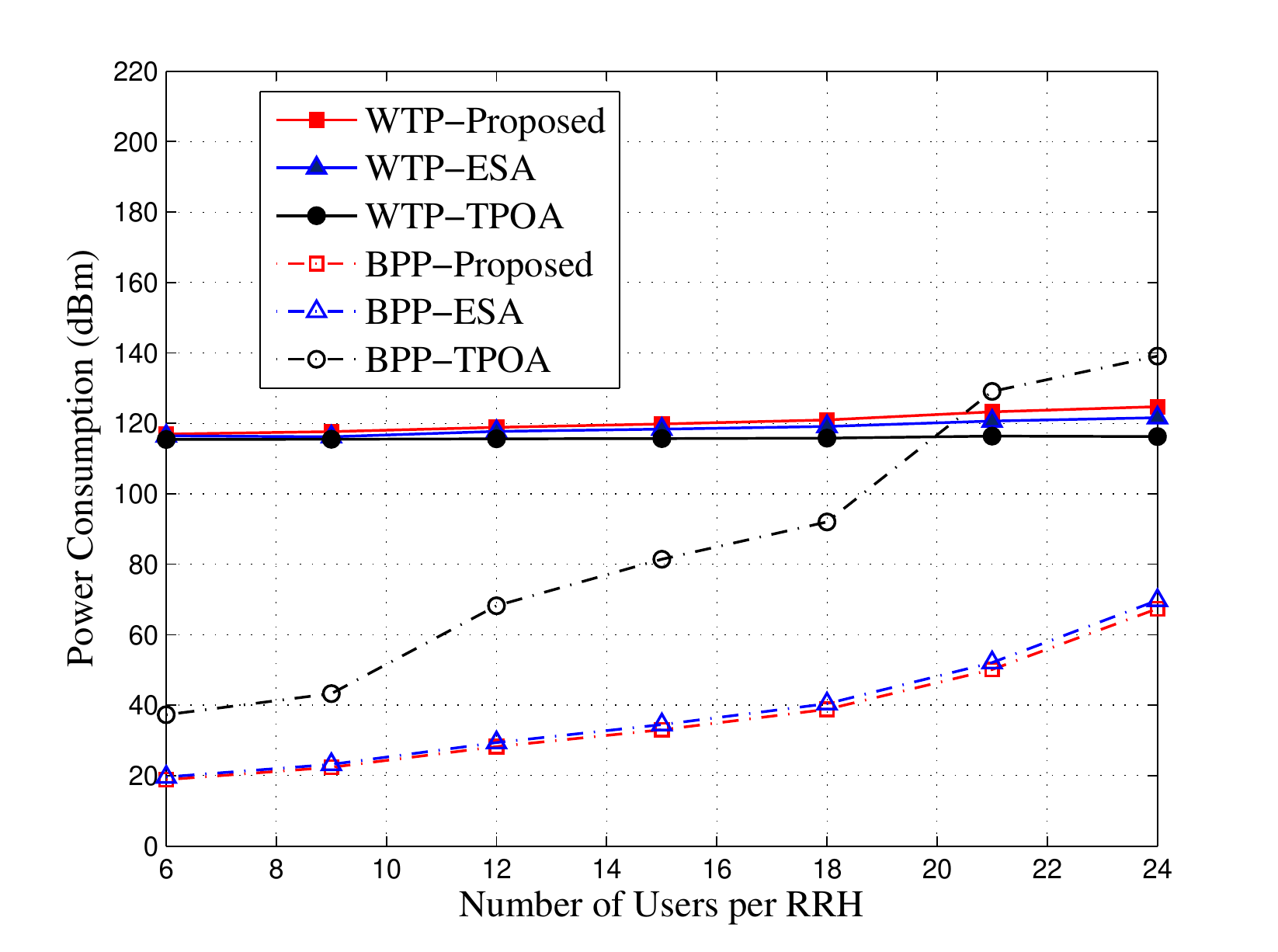}
  \caption{Power consumption vs. Number of users per RRH.}\label{fig6}
 \end{figure}

Fig. \ref{fig5} and Fig. \ref{fig6} show the power consumption varies with the number of users in each RRH. In the simulations, $9$ RRHs are deployed in the cellular network. The rate requirement of each user is set to $500$ Kbps. From Fig. \ref{fig5}, we can see that the proposed algorithm achieves similar performance as ESA and consumes much less power than TPOA where joint power optimization is not considered. From Fig. \ref{fig6}, we can see that the significant difference lies in the baseband processing power consumption for TPOA and the proposed algorithm. With TPOA, the baseband processing power consumption increases sharply when the number of users per RRH ranges from $9$ to $12$, and from $18$ to $21$. The reason is that a new BBU is activated in the BBU pool at these two points to serve more users. In terms of baseband processing power consumption, the proposed algorithm shows similar performance as ESA. When applying the proposed algorithm, we find the sharp increase only occurs when the number of users per RRH changes from $21$ to $24$. That is to say, the proposed algorithm needs less active BBUs than TPOA where joint power optimization is not considered. According to our analysis, load couples with wireless transmission power as well as baseband processing power. Compared with the proposed algorithm, TPOA results in more load and thus requires more active BBUs.

\section{CONCLUSIONS}

In this paper, we have studied load coupling power optimization in C-RAN. Both wireless transmission power and baseband processing power have been considered in the joint power optimization problem, which is based on load coupling model. We have shown that full load operation at each BS may not be optimal, which is different from the power optimization problem in the traditional BS system. To tackle the difficulties of solving such problem in C-RAN, we propose an efficient iterative algorithm. First, joint power optimization problem with a fixed number of active BBUs is solved. Then, the optimal power solution can be obtained by iteratively changing the number of active BBUs. Simulation results verify that the proposed algorithm outperforms existing algorithms where power optimization is not jointly considered. Furthermore, the performance of the proposed algorithm is quite close to that of the optimal algorithm with exhaustive search.

As future work, we will focus on joint optimization of spectral efficiency and energy efficiency in C-RAN based ultra-dense small cell networks.

\ifCLASSOPTIONcaptionsoff
  \newpage
\fi



%


\end{document}